\documentclass[preprint,preprintnumbers,aps,superscriptaddress,nofootinbib,tightenlines,floatfix]{revtex4}

\usepackage{amsmath,amssymb}
\usepackage{graphicx}
\usepackage{bm}
\usepackage{comment}
\usepackage{color}

\newcommand{\beq}{\begin{equation}}
\newcommand{\eeq}{\end{equation}}
\newcommand{\bea}{\begin{eqnarray}}
\newcommand{\eea}{\end{eqnarray}}

\newcommand{\Hbindabs}{ 13.2 \pm 1.8 \pm 4.0\mbox{ MeV} }
\newcommand{\HbindabsHALQCD}{ 37.4 \pm 4.4 \pm 7.3\mbox{ MeV} }
\newcommand{\HbindUsSecond}{ - 0.6 \pm 8.9 \pm 10.3\mbox{ MeV} }
\newcommand{\HbindQuad}{7.4\pm 2.1\pm 5.8\mbox{ MeV}}
\newcommand{\HbindLin}{-0.2\pm 3.3\pm 7.3\mbox{ MeV}}

\newcount\hour \newcount\hourminute \newcount\minute
\hour=\time \divide \hour by 60
\hourminute=\hour \multiply \hourminute by 60
\minute=\time \advance \minute by -\hourminute
\newcommand{\mydate}{\ \today \ - \number\hour :\number\minute}

\begin{document}

\preprint{
\vbox{
\hbox{UNH-11-1}
\hbox{ICCUB-11-125}
\hbox{JLAB-THY-11-1326}
\hbox{NT@UW-11-02}
\hbox{IUHET-558}
\hbox{UCB-NPAT-11-003}
\hbox{NT-LBNL-11-005}
}}

\begin{figure}[!t]
  \vskip -1.5cm
  \leftline{\includegraphics[width=0.25\textwidth]{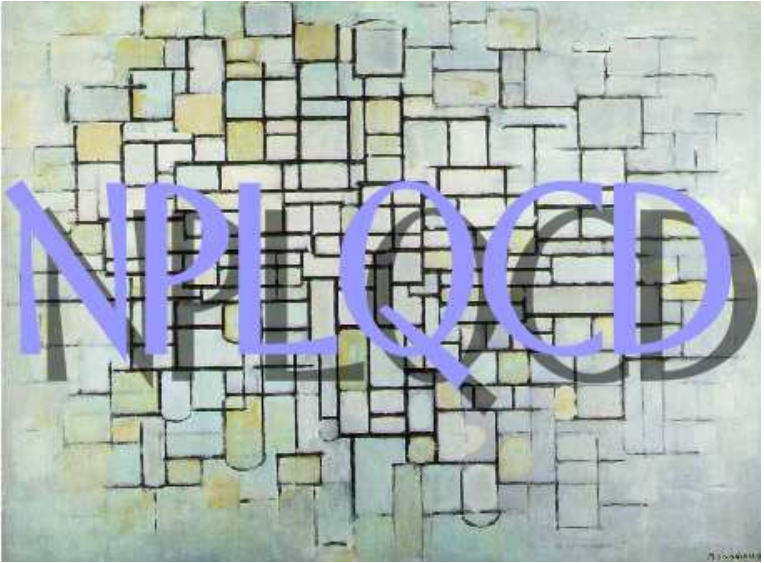}}
\end{figure}

\title{Present Constraints on the H-dibaryon at the Physical Point
from Lattice QCD}

\author{S.R.~Beane}
\affiliation{Albert Einstein Zentrum f\"ur Fundamentale Physik, Institut f\"ur theoretische Physik, Sidlerstrasse 5, CH-3012 Bern, Switzerland}
\affiliation{Department of Physics, University of New Hampshire, Durham, NH 03824-3568, USA}

\author{E.~Chang}
\affiliation{
Departament d'Estructura i Constituents de la Mat\`eria and
Institut de Ci\`encies del Cosmos,
Universitat de Barcelona, Mart\'{\i} i Franqu\`es 1, E08028-Spain}

\author{W.~Detmold}
\affiliation{Department of Physics, College of William and Mary, Williamsburg, VA 23187-8795, USA}
\affiliation{Jefferson Laboratory, 12000 Jefferson Avenue, Newport News, VA 23606, USA}

\author{B.~Jo\'o}
\affiliation{Jefferson Laboratory, 12000 Jefferson Avenue, Newport News, VA 23606, USA}

\author{H.-W.~Lin}
\affiliation{Department of Physics, University of Washington, Box 351560, Seattle, WA 98195, USA}

\author{T.C.~Luu}
\affiliation{N Division, Lawrence Livermore National Laboratory, Livermore, CA 94551, USA}

\author{K.~Orginos}
\affiliation{Department of Physics, College of William and Mary, Williamsburg, VA 23187-8795, USA}
\affiliation{Jefferson Laboratory, 12000 Jefferson Avenue, Newport News, VA 23606, USA}

\author{A.~Parre\~no}
\affiliation{
Departament d'Estructura i Constituents de la Mat\`eria and
Institut de Ci\`encies del Cosmos,
Universitat de Barcelona, Mart\'{\i} i Franqu\`es 1, E08028-Spain}

\author{M.J.~Savage} \affiliation{Department of Physics, University of Washington, Box 351560, Seattle, WA 98195, USA}

\author{A.~Torok} \affiliation{Department of Physics, Indiana University, Bloomington, IN 47405, USA}
\author{A.~Walker-Loud}
\affiliation{Lawrence Berkeley National Laboratory, Berkeley, CA 94720, USA}

\collaboration{ NPLQCD Collaboration }

\date{\mydate}

\begin{abstract}
\noindent
The current constraints from lattice QCD on the existence of the H-dibaryon are discussed. With only two significant lattice QCD calculations of the H-dibaryon binding energy at approximately the same lattice spacing, the forms of the chiral and continuum extrapolations to the physical point are not determined. In this brief report, we consider the constraints on the H-dibaryon imposed by two simple chiral extrapolations. In both instances, the extrapolation to the physical pion mass allows for a bound H-dibaryon or a near-threshold scattering state. Further lattice QCD calculations are required to clarify this situation.
\end{abstract}
\maketitle
\noindent
The possibility of a bound H-dibaryon (a six-quark state $uuddss$) was first proposed by Jaffe~\cite{Jaffe:1976yi} in 1977 using an MIT bag-model~\cite{Chodos:1974je} calculation which predicted that such a state would have a mass 81~MeV below the $\Lambda\Lambda$ threshold. Since then, experiments~\cite{Trattner:2006jn,Yoon:2007aq,Takahashi:2001nm,Yamamoto:2000wf} and other theoretical calculations~\cite{Sakai:1999qm,Mulders:1982da}, including a few pioneering lattice QCD calculations~\cite{Mackenzie:1985vv,Iwasaki:1987db,Pochinsky:1998zi,Wetzorke:1999rt,Wetzorke:2002mx,Luo:2007zzb}, have been seeking evidence of its existence.
In 1997, Bashinsky and Jaffe~\cite{Bashinsky:1997qv} reviewed 28 model predictions on the mass of the H-dibaryon, which vary from 1.1 to 2.9~GeV. (See Fig.~4 in Ref.~\cite{Bashinsky:1997qv} for a histogram of model predictions.)
Two years later, Sakai et~al.~\cite{Sakai:1999qm} comprehensively reviewed various updated experimental and theoretical results (see Fig.~1 in Ref.~\cite{Sakai:1999qm}), extending the upper limit of model predictions to around 3.1~GeV. Model predictions for the H-dibaryon mass have not changed much since then.
Recently, the NPLQCD and HALQCD collaborations reported results showing that the H-dibaryon is bound for a range of light-quark masses larger than those found in nature~\cite{Beane:2010hg,Inoue:2010es}. These calculations are important for a number of reasons. First, they show that lattice QCD is now capable of calculating the energy of simple nuclei, with the H-dibaryon being an exotic example. Second, they provide evidence that a bound H-dibaryon may exist for some values of parameters entering the QCD Lagrangian.  However, it is important to determine if this system is, in fact, bound at the physical values of
the light-quark masses and with the inclusion of electroweak interactions.
Experimental evidence currently suggests that such a bound state does not exist
between 2.136 and 2.231~GeV~\cite{Trattner:2006jn} (indicating either that the binding energy could be larger than 95~MeV or there is no bound $H$ state at all),
but that a near-threshold resonance may exist in the scattering channel with the quantum numbers of the H-dibaryon~\cite{Yoon:2007aq}.
In this note we investigate the current constraints on the binding energy of the H-dibaryon at the physical value of the pion mass, in the isospin limit and in the absence of electroweak interactions, by extrapolating the available lattice QCD results.

Early quenched lattice QCD calculations~\cite{Mackenzie:1985vv,Iwasaki:1987db,Pochinsky:1998zi,Wetzorke:1999rt,Wetzorke:2002mx,Luo:2007zzb}
drew inconsistent conclusions regarding the binding of the H-dibaryon. Table~\ref{tab:SP_summary} shows a summary of parameters used in those works and their conclusions.
Generally the calculations involved less than 100 measurements, and the analyses lacked estimates of systematic uncertainties, such as ${\cal O}(b)$ discretization errors for Wilson fermions, finite-volume effects, and quenching (could be 30\% or larger depending on the physical quantity).

\begin{table}[!h]
\begin{center}
\caption{\label{tab:SP_summary}Summary of  H-dibaryon lattice QCD calculations.
The pion masses used in Refs.~\cite{Pochinsky:1998zi,Wetzorke:1999rt,Wetzorke:2002mx} are estimated from the bare quark-mass parameters listed in their proceedings.
For the works that do not perform volume extrapolation to the $V \rightarrow \infty$ limit, the mass difference between the H-dibaryon and 2-$\Lambda$ state
from the largest volume is used. }
\footnotesize
\begin{tabular*}{160mm}{c@{\extracolsep{\fill}}|c|c|c|c|c|c|c}
\hline\hline
Ref. & $n_f$ & $S_f$ & $a_t^{-1}$ (GeV)  & $m_\pi$ (GeV) & $L$ (fm) & $V \rightarrow \infty $  & $m_H-2m_\Lambda$ (MeV) \\
\hline
Mackenzie et~al.~\cite{Mackenzie:1985vv} & 0 & Wilson & 0.9 & 0.60--0.90 & 1.3 & no & $>0$ \\
Iwasaki et~al.~\cite{Iwasaki:1987db}    & 0   & Wilson & 1.81   & 0.70--1.90 &  1.74  & no & [$-676$,$-381$] \\
Pochinsky et~al.~\cite{Pochinsky:1998zi}    & 0   & Wilson & 1.79  & $>0.5$  &  2.08, 3.12  & yes & $>0$ \\
Wetzorke et~al.~\cite{Wetzorke:1999rt,Wetzorke:2002mx}    & 0   & clover & 1.11   & $>0.5$ &  1.4--4.27  & no & $>0$  \\
Luo et~al.~\cite{Luo:2007zzb} & 0   & aClover* & 1.48 & 0.70--0.85 & 3.2,6.4 & no & $-64(59)$ \\
\hline
NPLQCD (this work)  & 2+1   & aClover* & 5.6   & 0.39 &  2.0, 2.5, 3.0, 4.0 & yes & $-13.2(1.8)(4.0)$ \\
NPLQCD (this work)    & 2+1   & aClover* & 5.6   & 0.23 & 4.0 & no & $0.6(8.9)(10.3)$ \\
HALQCD~\cite{Inoue:2010es}    & 3   & clover & 1.63   & 0.67--1.01 & 2.0, 3.0, 4.0  & yes & $[-40,-30]$ \\
\hline\hline
\end{tabular*}
\end{center}
{\footnotesize * The abbreviation ``aClover'' stands for the anisotropic clover fermion action.}
\end{table}

Unlike the single-hadron spectroscopy of a typical lattice QCD calculation,
high statistics are essential to manage the uncertainties in the extraction of bound-state information. To demonstrate this, we have broken down our measurements into smaller subsets to see how the uncertainty evolves~\cite{Beane:2009py}.
Figure~28 in Ref.~\cite{Beane:2009py} shows NPLQCD's extracted values of $(k\cot\delta)^{-1}$ for $\Lambda\Lambda$ scattering ($\delta$: phase shift) versus the extracted value of $|{\bf k}|^2$ ($\bf k$: center-of-mass momentum) with total number of measurements varying from  $10^4$ to $4\times10^5$
 in a lattice with spatial extent of 2.5~fm.
With a small number of measurements, one could easily extract a value of $k\cot\delta$ that appears repulsive at this particular volume
without a careful estimation of the systematic uncertainty.
Even with $10^5$ total measurements, we are not yet able to conclusively determine whether the interaction is attractive or repulsive.
It is only when the total number of measurements approaches $4\times10^5$ that the interaction can be determined to be attractive, and even then with only about $2\sigma$ significance.
For the remainder of this brief report, we will concentrate on the two recent dynamical studies with higher statistics by the NPLQCD and HALQCD collaborations.

The details of the two lattice QCD calculations that provide
statistically significant evidence for a bound H-dibaryon can be found
in the recent works of NPLQCD~\cite{Beane:2010hg} and
HALQCD~\cite{Inoue:2010es}.  The NPLQCD result is determined from
$n_f=2+1$
calculations
in four lattice volumes (with spatial extents
of $L\sim 2.0, 2.5, 3.0$ and $4.0$~fm), each at a single spatial
lattice-spacing of $b\sim 0.123$~fm and a pion mass of
$m_\pi\sim 390$~MeV.
Ref.~\cite{Beane:2010hg} reported a binding energy of $B_H={ 16.6 \pm 2.1 \pm 4.6\mbox{ MeV} }$ at that pion mass; with more statistics on this ensemble, the updated binding energy is now $\Hbindabs$.
The HALQCD collaboration performed
calculations in three lattice volumes (with spatial extents
of $L\sim 2.0, 3.0$ and $4.0$~fm) at a lattice spacing of $b\sim
0.121$~fm and in the limit of SU(3) flavor symmetry ($n_f=3$)
at three
different quark masses giving $m_\pi\sim 673, 837$ and
$1015$~MeV.  In order to extrapolate in the
light-quark masses, the binding energy of $B_H=\HbindabsHALQCD$ obtained at
$m_\pi\sim 837$~MeV is used because this pion mass corresponds
to a strange-quark mass that is closest to that of nature\footnote{
At leading order in chiral perturbation theory $m^2_\pi = 2 \lambda \overline{m}$
and $m^2_K =  \lambda (m_s + \overline{m})$ (Gell-Mann--Oakes--Renner)
in the limit of isospin symmetry,
where $\overline{m}$ is the isospin-averaged light quark mass and $\lambda$ is
related to the quark condensate.
Inserting the NPLQCD values of the pion and kaon masses into this relation
would lead to using the H-dibaryon binding energy calculated
at $m_\pi\sim 673$~MeV by HALQCD.
However, lattice QCD calculations
(in particular, calculations on
additional ensembles that  have the same strange quark mass as those
used in the  NPLQCD calculations,
as shown in Table VI of Ref.~\cite{Lin:2008pr}), which include the
full light-quark mass dependence, dictate that the binding energy calculated at
$m_\pi\sim 837$~MeV is the appropriate value to use for the SU(3)
symmetric point.  Given the size of the uncertainties of both lattice QCD calculations, the difference in the results obtained with the
$m_\pi\sim 837$~MeV and $m_\pi\sim 673$~MeV are not
statistically significant.
}
(and to that of
the NPLQCD calculations)\footnote{
In this channel,
the ground-state energy-levels
determined in the lattice QCD calculations of HALQCD~\cite{Inoue:2010es}
are expected to be exponentially close to the actual bound-state energy of the
H-dibaryon and do not suffer from the uncontrolled approximations
that are present~\cite{Beane:2010em} in phase-shifts calculated via
the energy-dependent and sink-dependent potentials presented by HALQCD.}.

NPLQCD and HALQCD employed different clover
discretizations of the light quarks (and different gauge-actions),
providing results that are ${\cal
  O}(b)$-improved and therefore both sets of calculations have
lattice-spacing artifacts that scale as ${\cal O}(b^2)$.
Given the precision with which the single-hadron
energy-momentum relation is satisfied~\cite{Beane:2010hg}, the
contributions from
Lorentz-symmetry breaking operators that appear at order ${\cal O}(b^2)$ are
expected to be highly suppressed.
Naive scaling arguments, as well as the cancellations that
occur in forming energy differences, suggest that lattice spacing
artifacts are suppressed, as compared, for instance, to the leading
quark-mass effects.  However, definitive statements about the lattice spacing dependence
will require calculations at smaller lattice spacings. In this context,
we assume that lattice spacing artifacts are small (and therefore
are not driving the H-dibaryon binding), and focus on the light-quark mass
dependence.
Alternately, one can view the present study
as performing the chiral extrapolation of the
(approximately) fixed lattice-spacing results.

The form of the chiral extrapolation of the H-dibaryon binding energy
is unknown, and there are a number of features that make it difficult
to determine.  In the limit of exact SU(3) flavor symmetry, the
non-interacting ground state of the $I=0$, $J=0$ and $s=-2$ (the
quantum numbers of $\Lambda\Lambda$) system is multiply degenerate,
comprised of the hadronic states $\Lambda\Lambda$, $\Sigma\Sigma$, and
$N\Xi$.  The interactions among these states produce the H-dibaryon
(when it is bound) as the ground-state, with the other states
being of higher energy.  Even in the presence of SU(3) breaking, as is
the case in the NPLQCD calculations, the spectrum is qualitatively
similar with the non-interacting states being nearly degenerate.  If
the H-dibaryon is tightly bound, then a chiral
expansion\footnote{Low-energy effective field theories have been
constructed to describe baryon systems carrying strangeness;
see Refs.~\cite{Savage:1995kv,Beane:2003yx,Polinder:2006zh,Haidenbauer:2009qn}.
}
of the form
of that for single hadrons will result.
On the other hand, if the H-dibaryon is loosely bound,
the binding energy results from a cancellation between short-distance
and long-distance contributions, as is the case with the
deuteron where the extrapolation has significant
structure~\cite{Beane:2002vs,Beane:2002xf,Epelbaum:2002gb,Mondejar:2006yu}.
However, in contrast with the NN system,
the long-distance contribution
to $\Lambda\Lambda$ interactions
results from two-pion exchange.
Consequently,
the H-dibaryon is expected to be less fine tuned than the deuteron,
leading to the
expectation that the chiral extrapolation may resemble
the form $B_H(m_\pi) = B_0 + d_1 m_\pi^2 + {\cal O}(m_\pi^3)$.

The quark mass dependence of the lowest-lying octet baryon masses, such as
the N, $\Lambda$, $\Sigma$ and $\Xi$, is an ongoing topic of discussion,
and motivates a second extrapolation form.
The form of the light-quark mass dependence of the baryon masses
produced by the lowest orders of baryon chiral perturbation theory does
not naturally reproduce the currently available results of lattice QCD calculations\footnote{SU(2) heavy-baryon chiral perturbation theory
does reproduce the lattice
results for the nucleon mass (with $\chi^2/{\rm dof}\sim 1$),
but with large cancellations between different orders in the expansion~\cite{WalkerLoud:2008bp}.}.
This is currently interpreted as an indication
that the chiral expansion is not converging at these heavier pion masses and the possibility
there is a new
scale associated with the strong interactions~\cite{Hall:2011jr}.  At
the least, there is correlation between higher orders in the expansion
beyond the expectations from naive dimensional analysis.
Walker-Loud~\cite{WalkerLoud:2008pj} performed a comprehensive
analysis of all of the lattice QCD calculations of the nucleon mass
and found that the results are consistent with linear dependence on
the dimensionless variable $m_\pi/f^{(0)}_\pi$, where $f^{(0)}_\pi$ is the pion
decay constant in the chiral limit, with an extrapolation that is
consistent with the experimental value.  This is in
contrast with the expectations of an expansion about the chiral limit,
which has the form $M_N(m_\pi) = M_0 \ +\ \alpha_2 m_\pi^2\ +\ {\cal
  O}(m_\pi^3)$, where $M_0$ and $\alpha_2$ are parameters that must be
determined from the lattice QCD calculations.
This motivates us to consider a non-analytic  extrapolation of the
form $B_H(m_\pi) = \tilde B_0 + c_1 m_\pi + {\cal O}(m_\pi^2)$.
It is possible that the true form lies somewhere between
the linear and the quadratic forms, with cancellations occurring between
analytic and non-analytic contributions.  Without any better guidance
as to the form of the chiral extrapolation, we will consider
the results from these two forms of extrapolation with relatively heavy
pions to provide nothing more than an estimate of the H-dibaryon
binding energy at the physical light-quark masses.
\begin{figure}[!h]
  \centering
  \includegraphics[width=0.48\columnwidth]{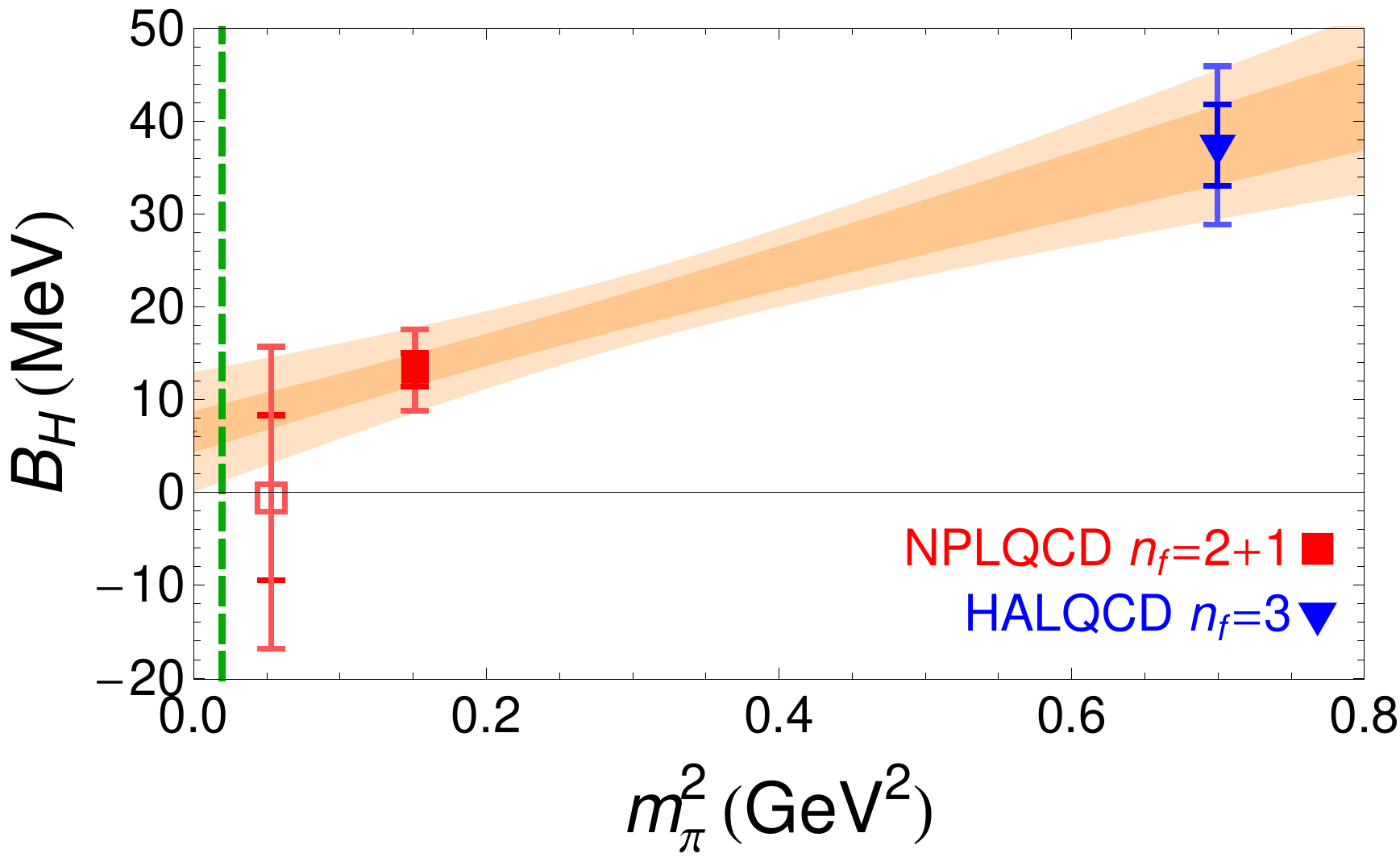}
  \includegraphics[width=0.48\columnwidth]{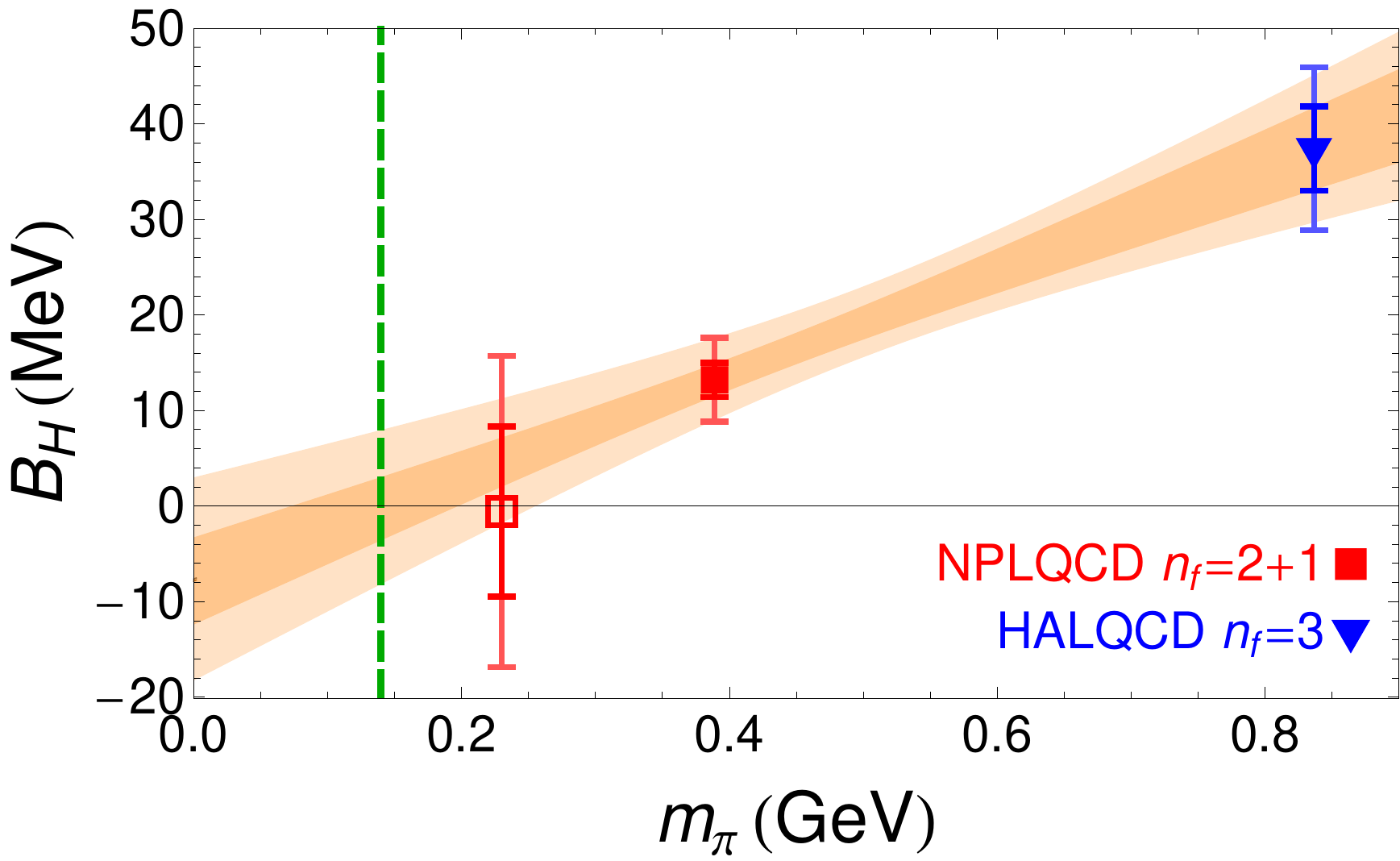}
  \caption{
The results of lattice QCD from the NPLQCD
collaboration~\cite{Beane:2010hg}  and this work
($n_f=2+1$) (red squares), and the
 HALQCD collaboration~\cite{Inoue:2010es} ($n_f=3$) (blue triangles).
The filled symbols are used in the extrapolations, while the open squares (NPLQCD's 230-MeV data) are not.
    Left panel:
    The darker (lighter) shaded region corresponds to an extrapolation
    of the LQCD calculations that is quadratic in the pion mass, of
    the form $B_H(m_\pi) = B_0\ +\ d_1\ m_\pi^2$ where the parameters are
    determined by the central values and statistical uncertainties
    (statistical and systematic uncertainties combined in quadrature).
    The vertical dashed (green) line corresponds to the physical pion
    mass.
    Right panel: Same as the left panel but with the extrapolation
$B_H(m_\pi) = \tilde B_0\ +\ c_1\ m_\pi$.
  }
  \label{fig:ExtrapsQuadLinear}
\end{figure}
Extrapolations with more complicated behaviors are allowed but cannot be
constrained by the current lattice QCD calculations, and are not
discussed further.
The chiral extrapolation of the H-dibaryon binding energy using the
form $B_H(m_\pi) = B_0\ +\ d_1\ m_\pi^2$ results in the shaded region
shown in Fig.~\ref{fig:ExtrapsQuadLinear} (left panel).
The H-dibaryon binding
energy at the physical value of the pion mass,
neglecting isospin-violation and electromagnetic interactions,
is found to be
\begin{eqnarray}
B_H^{\rm quadratic} & = & \HbindQuad
\ \ \ ,
\label{eqnquadraticextrap}
\end{eqnarray}
as indicated by the intercept of the shaded region with the (green)
dashed line in Fig.~\ref{fig:ExtrapsQuadLinear} (left panel).
The first uncertainty results from an extrapolation using the statistical
uncertainties of both lattice QCD calculations, while the second uncertainty results
from the systematic uncertainties.
The quadratic extrapolation
suggests that the H-dibaryon is bound at the physical value of the
pion mass.  However, the
H-dibaryon is unbound at the $2\sigma$ level, and a near threshold
scattering state remains allowed by the current lattice QCD calculations.
Further, at the  $2\sigma$ level, the extrapolation is also consistent with
the binding energy
being
independent of $m_\pi$.

Using the form $B_H(m_\pi) = \tilde B_0\ +\ c_1\ m_\pi$ to
chirally extrapolate the  lattice QCD calculations produces the results
shown in Fig.~\ref{fig:ExtrapsQuadLinear} (right panel).  The
H-dibaryon binding energy at the physical value of the pion mass is
found to be
\begin{eqnarray}
B_H^{\rm linear} & = & \HbindLin
\ \ \ ,
\label{eqnlinearextrap}
\end{eqnarray}
as indicated by the intercept of the shaded region with the (green) dashed line
in Fig.~\ref{fig:ExtrapsQuadLinear} (right panel).
With the precision of the current lattice QCD results,
the linear chiral extrapolation does not discriminate between a bound or
unbound  H-dibaryon  at the physical
pion mass.
However, it suggests that  there is a state in the
 $I=0$, $J=0$ and $s=-2$ channel that is either just bound or just unbound.

Currently, both NPLQCD and HALQCD are improving their calculations to include lighter pion masses.
NPLQCD is working on a calculation using an $n_f=2+1$ $m_\pi\sim 230$~MeV ensemble, which at sufficiently high statistics would 
improve our ability to extrapolate the H-dibaryon binding energy to the physical limit and determine whether it remains bound.
To date, a single lattice volume with spatial extent 4~fm (giving $m_\pi L$ around 4.7) has been completed with around $9\times 10^4$
measurements performed, and a binding energy of $B_H=\HbindUsSecond$ is obtained.
This point is included for illustrative purposes in Fig.~\ref{fig:ExtrapsQuadLinear}, shown as an open square in both figures.
However, an infinite volume extrapolation is currently unavailable at this pion mass, and we have not estimated the associated systematic uncertainty.
Unfortunately, at this stage, this additional information does not differentiate between the extrapolation forms, and
further improvements are needed before this light pion-mass ensemble can provide significant input to the extrapolation.

We conclude that the current lattice QCD calculations of
NPLQCD and HALQCD
chirally extrapolated with plausible forms for the light-quark mass
dependence, are not sufficiently precise at light enough quark masses
to determine whether QCD predicts a
 bound H-dibaryon.
However, the current lattice QCD  calculations hint that the H-dibaryon becomes less bound as the
 light-quarks become lighter, and the extrapolations suggest that
 there is a near-threshold state in the spectrum.
Calculations at smaller lattice spacings are also required to ensure that
 lattice artifacts are not driving the observed H-dibaryon
binding, and to allow extrapolation to the continuum limit.

This result should be viewed in the context of the present experimental constraints, which, at face-value, effectively eliminate the possibility of a bound H-dibaryon~\cite{Trattner:2006jn}\footnote{The constraints presented in Ref.~\cite{Trattner:2006jn} are model-dependent and may be weaker than claimed. In particular, if the H-dibaryon were a compact object on the scale of two baryons, then the present experimental bounds would be evaded~\cite{Farrar:2003qy}.}.
However the suggestion of structure in the scattering amplitude near threshold~\cite{Yoon:2007aq} would not be inconsistent with the current lattice QCD results.
In order to refine the lattice-QCD prediction, precision calculations at lighter quark masses are required.
We are currently increasing the statistics of the calculations on the 230-MeV ensemble; this would significantly improve our ability to predict whether the H-dibaryon
is bound or unbound at the physical light-quark masses.
A multi--lattice-spacing, multi-volume lattice QCD study at the physical pion mass would be unambiguous, up
to systematic uncertainties that arise from working in the isospin limit in the absence of electromagnetism.
Ultimately, these effects will also be included.

\vskip 0.2in

We would like to thank R. Jaffe for prompting us to consider the
chiral-extrapolation, and D. Hertzog for
emphasizing the importance of performing the extrapolation.

\vskip 0.2in

SRB was supported in part by the NSF CAREER grant PHY-0645570.  The Albert Einstein Center for Fundamental Physics is supported by the “Innovations- und Kooperationsprojekt
 C-13” of the “Schweizerische Universit\"atskonferenz SUK/CRUS”.  The
 work of EC and AP is supported by the contract FIS2008-01661 from MEC
 (Spain) and FEDER.  AP acknowledges support from the RTN Flavianet
 MRTN-CT-2006-035482 (EU).  H-WL and MJS were supported in part by the
 DOE grant DE-FG03-97ER4014.  WD, KO and BJ were supported in part by DOE
 grants DE-AC05-06OR23177 (JSA) and DE-FG02-04ER41302.  WD was also
 supported by DOE OJI grant DE-SC0001784 and Jeffress Memorial Trust,
 grant J-968.
BJ was also supported by DOE grants, DE-FC02-06ER41440 and DE-FC02-06ER41449
(SciDAC USQCD)
KO was also supported in part by NSF grant CCF-0728915,
Jeffress Memorial Trust grant J-813 and DOE OJI grant
DE-FG02-07ER41527.  AT was supported by NSF grant PHY-0555234 and DOE
grant DE-FC02-06ER41443.  The work of TL was performed under the
auspices of the U.S.~Department of Energy by LLNL under Contract
DE-AC52-07NA27344.
The work of AWL was supported in part by the Director, Office of Energy
Research, Office of High Energy and Nuclear Physics, Divisions of
Nuclear Physics, of the U.S. DOE under Contract No.  DE-AC02-05CH11231


\end{document}